\documentclass[aps,prl,amsmath,twocolumn,showpacs,amssymb,superscriptaddress]{revtex4}
\usepackage{graphicx} 
\usepackage{color}
\usepackage{times}
\usepackage{tabularx}
\usepackage{bm}
\begin{document}
\title{Quantum communication protocols by quantum walks with two coins}

\author{Yun Shang}
\email{corresponding author: shangyun602@163.com}
\affiliation{NCMIS, Academy of Mathematics and Systems Science,
   Chinese Academy of Sciences, Beijing, 100190, China}
\affiliation{MDIS, Academy of Mathematics and Systems Science,
   Chinese Academy of Sciences, Beijing, 100190, China}
\author{Yu Wang}
\affiliation{Academy of Mathematics and Systems Science,
   Chinese Academy of Sciences, Beijing,100190, China}
\affiliation{University of Chinese Academy of Sciences, Beijing, 100049, China}
\author{Meng Li}
\affiliation{Academy of Mathematics and Systems Science,
   Chinese Academy of Sciences, Beijing,100190, China}
\affiliation{University of Chinese Academy of Sciences, Beijing, 100049, China}

\author{Ruqian Lu}
\affiliation{NCMIS, Academy of Mathematics and Systems Science,
   Chinese Academy of Sciences, Beijing, 100190, China}
\affiliation{MDIS, Academy of Mathematics and Systems Science,
   Chinese Academy of Sciences, Beijing, 100190, China}
\affiliation{ Key Lab of IIP, Institute of Computing Technology, CAS, Beijing, 100190, China}

\date{\today}
\begin{abstract}

 We introduce some new perfect state transfer and teleportation schemes by quantum walks with two coins. Encoding the transferred information in coin 1 state and alternatively using two coin operators, we can perfectly recover the information on coin 1 state at target position only by at most two times of flipping operation. Based on quantum walks with two coins either on the line or on the $N$-circle, we can perfectly transfer any qubit state. In addition, using quantum walks with two coins either on complete graphs or regular graphs, we can first implement perfect qudit state transfer by quantum walks. Compared with existing schemes driven by one coin, more general graph structures can be used to perfectly transfer more general state. Moreover, the external control of coin operator during the transmitting process can be decreased greatly. Selecting coin 1 as the sender and coin 2 as the receiver, we also study how to realize generalized teleportation over long steps of walks by the above quantum walk models. Because quantum walks is an universal quantum computation model, our schemes may provide an universal platform for the design of quantum network and quantum computer.

\end{abstract}

\maketitle

\section{I. Introduction}

Quantum walks have been introduced as a quantum analogue of classical random walks. It was first proposed by Aharonov \emph{et al.} in 1993 \cite{Aharonov_1993}, and the model of quantum walks on the line was developed  by Ambainis \emph{et al.} in 2001 \cite{Ambainis_2001}. For the case of a general graph, Aharonov \emph{et al.} \cite{Aharonov_2001} also introduced the corresponding model. After these works, there appeared much fruitful research on discrete and continuous quantum walks \cite{Kempe_2003,Venegas_2012}. One kind of the prominent applications of quantum walks is to design algorithms in quantum information processing, such as search algorithms, and element distinctness, etc. \cite{Ambainis_2001,Buhrman_2001,Kempe_2003,Ambainis_2004,Szegedy_2004,Shenvi_2003,Ambainis_2007,Childs_2010}. Another kind of applications for quantum walks is chosen to be an universal quantum computation model \cite{Lovett_2010,Childs_2009,Childs_2013}.
Quantum walks with multiple coins was first proposed by Brun \emph{et al.} in 2003 \cite{Brun_2003}. However, because it has some similar properties with quantum walks with one coin, it hasn't drawn much attention. Recently, we found that it can finish some quantum communication tasks more conveniently \cite{Wang_2017}. In this paper, we mainly consider quantum state transfer and quantum teleportation protocols.

As far as we know, quantum state transfer between different sites is a significant problem for quantum networks and quantum computer. In 2003, Bose first considered this problem using spin chains
as the quantum communication carrier in quantum computing \cite{Bose_2003}. Because spin chain can be regarded as a wire in quantum networks, state transfer by spin chains has been studied widely \cite{Christandle_2004,Di_2008,Nikolopoulos_2014}. An ideal scenario is that the unit fidelity of transfer is needed. However, it has been demonstrated that such perfect transfer cannot be achieved within chains with basic nearest-neighbor couplings \cite{Nikolopoulos_2014}.

 In 2004, Chiristandle\emph{ et. al} found that the time-evolution of qubit state transfer through the spin chain can be interpreted as a continuous time quantum walk \cite{Christandle_2004}. It was also generalized to discrete time quantum walk by adding a position-dependent coin operator \cite{Kurzynski_2011}. In the last decade, state transfer by quantum walks became an interesting topic. Kendon \emph{et al.} studied state transfer in quantum walks on various graphs, and they mainly considered continuous time quantum walks. They found that very few graphs can realize perfect state transfer \cite{Kendon_2010, Barr_2013}. By discrete-time quantum walk search algorithm,
M. Stefa\v{n}\`{a}k \emph{et al.} discussed state transfer on star graphs and complete bipartite graphs \cite{Stefanak_2016}. Yalcinkaya \emph{et al.} \cite{Yalcinkaya_2015} designed qubit state transfer with discrete-time quantum walk on a $N$-circle and $N$-line by adding some recovery operator to the whole system. However, for the $N$-circle protocol, $N$ should be even number, and the transferred state can only be transmitted to the opposite site on the circle. Zhan \emph{et al.} \cite{Zhan_2014} designed paths to transfer qubit state by discrete time quantum walks on the line. However, at the same step, different coin operators is needed to do flipping operation for different positions. Obviously, they will bring extra complexity for operation in reality experiment.

In order to implement scalable quantum networks, we need to set up various building blocks of the networks such as line, circle, complete graph or regular graph. To observe or control corresponding quantum system, one has to select an interface to interact with the system. The interface can also be used as a quantum communication channel \cite{Lloyd_2004}. Here selecting quantum walks with two coins as the basic model and then two coin space as the communication carrier, we successfully implement some quantum communication tasks on the above graphs.

In this paper, we consider both state transfer and generalized teleportation schemes driven by quantum walks with two coins. We select coin spaces as the quantum communication channel. In the first scheme, we encode the unknown state at the coin 1 space. After finite steps of walk, by alternatively using the two coin operators, we can transfer the state to any target position, and recover transferred information perfectly in coin 1 space at the target position. Based on this idea, we find that any unknown qubit state can be perfectly transferred on the line and on the $N$-circle by using these two coins alternatively. Compared with results given by Yalcinkaya \cite{Yalcinkaya_2015}, we can transfer information to any target position in an infinite line. As for $N$-circle, our circle is not necessarily to have even vertex, and the information can be transferred from the initial position to any other position on the circle. Compared with routing path on the line \cite{Zhan_2014}, for the same step, we use the uniform coin operator at different positions. Obviously, it reduces the extra complexity for operating in practice experiment. In our scheme, only at most two moments, we need to use coin flipping operator Pauli $X$, and we can calculate the moment accurately. However, more than two moments is needed to do coin flipping operation \cite{Zhan_2014}. Most importantly, we first show that an unknown qudit state can be perfectly transferred on the complete graph or regular graph by quantum walks.

The second scheme is to implement generalized teleportation over long distance walks by two coins. This is a further extension of the work \cite{Wang_2017}. By two steps of quantum walk with two coins, we can transmit qubit state on the line or on the 4-circle. Moreover, we can teleport any qudit state on either completer graph or regular graph \cite{Wang_2017}. Although one can use classical teleportation scheme successfully in experiment when teleporting a lower dimension state \cite{Bennett_1993}, how to teleport any qudit state is still a very difficult problem. However, for $d > 2$, our schemes embody its advantage. By changing the $d^{2}$-dimensional measurement basis of the original protocol into two $d$-dimensional measurements, quantum walk formalism might make it possible to have a simpler implementation than the original teleportation protocol. In \cite{Wang_2017}, we implement teleportation only by two steps of walk. In this paper, we study how to realize teleportation by quantum walks with two coins after long steps of walk. Encoding the unknown information on coin 1 space, after long distance of walks, we can still recover the information on coin 2 space perfectly by local measurement and local correction. Based on our schemes, we can implement unknown qubit teleportation over long distance walks on the circle with even vertex. After $2t$ steps of walk, we can implement any qudit state teleportation using complete graph or regular graph.

Because quantum walks is an universal quantum computation model, these communication protocols may provide a universal platform for the design for quantum computer.

The paper is organized as follows. In Section II, we introduce models of coined quantum walks on various graphs. In Section III, various state transfer schemes based on quantum walks with two coins are given. In Section IV, long distance teleportation by quantum walks with two coins on various graphs are showed. In the last section, we give the conclusion.

\section{II. Prelimanary}
\textbf{Quantum walks on the line}
One-dimensional quantum walk \cite{Ambainis_2001} takes place in a compound Hilbert space comprising position space and coin space, defined as $\mathcal{H}=\mathcal{H}_{P}\otimes\mathcal{H}_{C}$, where $\mathcal{H}_{P}=\rm{span}\{|n\rangle:n\in Z\}$ and $\mathcal{H}_{C}=\rm{span}\{|0\rangle,|1\rangle\}$. The walker walks in a one-dimensional line. Each step of the quantum walk is described by the equation $W^{(l)}=S\cdot(I\otimes C)$, where $C$ is called a coin operator acting on coin space, and the conditional shift operator $S$ is denoted by
\begin{equation}
S=\sum_{x}(|x+1\rangle\langle x|\otimes|0\rangle\langle0|+|x-1\rangle\langle x|\otimes|1\rangle\langle1|).
\end{equation}
The conditional shift operator simulates the classical way of random walks. The walker moves either one step to the left or to the right depending on the state of the coin.

\textbf{Quantum walks on graphs}
  The definition of coined quantum walks on graphs was given by Aharonov \cite{Aharonov_2001}. Let $G(V,E)$ be a graph, and $\mathcal{H}_{v}$ be the Hilbert space spanned by states $|v\rangle$, where vertex $v\in V$. At each vertex $j$, there are several directed edges with labels pointing to the other vertices. The coin space $\mathcal{H}_c$ is spanned by states $|a\rangle$, where $a\in\{0,\cdots,d-1\}$. They are the labels of the directed edges. The conditional shift operation between $\mathcal{H}_v$ and $\mathcal{H}_c$ is $T=\sum_{j,a}|k\rangle\langle j|\otimes|a\rangle\langle a|$, where the label $a$ directs the edge $j$ to $k$.

  \textbf{Quantum walks driven by many coins}

\begin{figure}[ht]
\centering
\includegraphics[width=6cm]{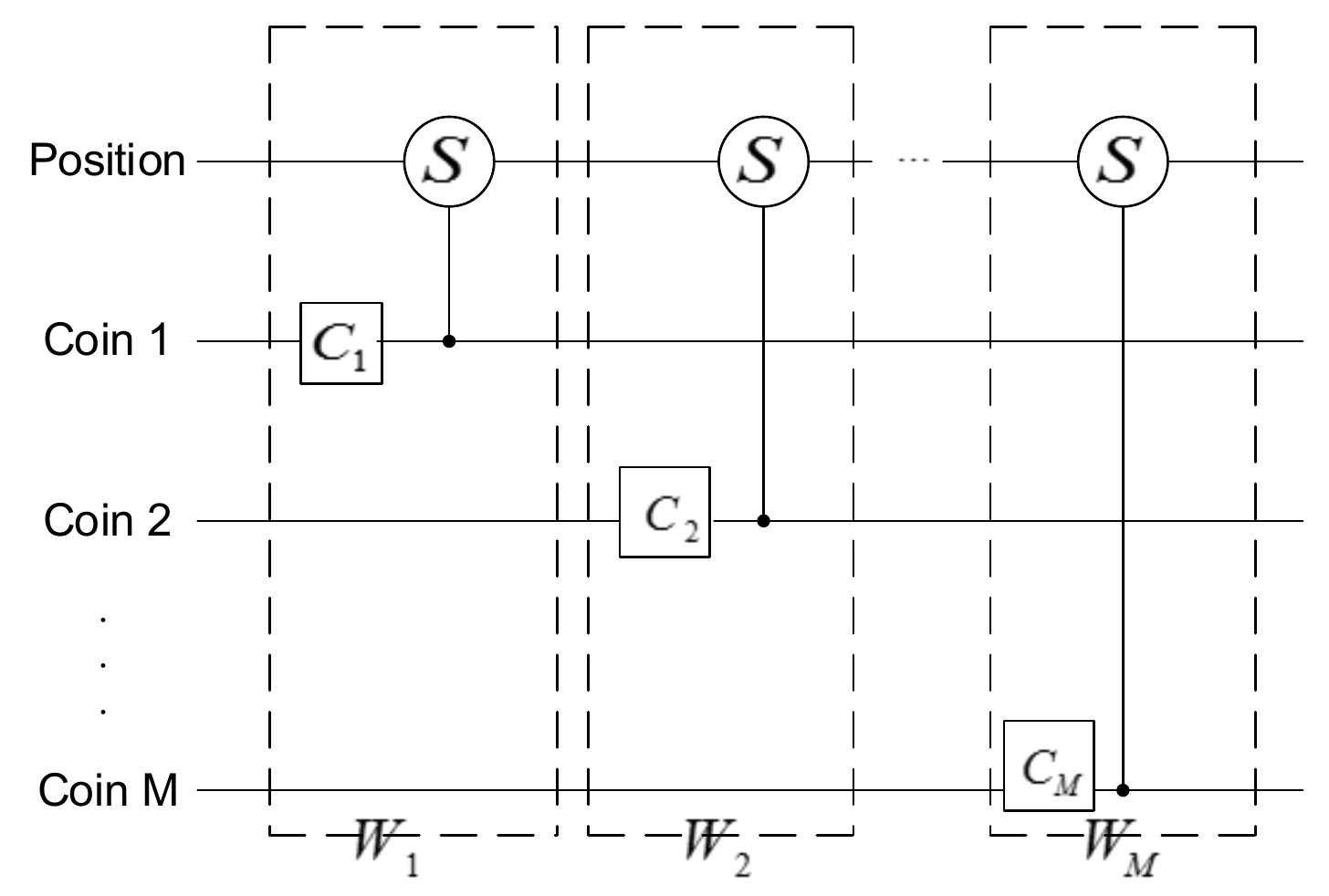}
\caption{Circuit diagram of quantum walks with multiple coins.  The first horizontal line represents the walker qudit (a superposition of finite many positions), and the following lines are coin states. $C_{i}$ is a coin operator on the $i$th coin. $S$ denotes the transition operator on the position space. The control state is in the coin's space. The target state is in the position space. When the control state is $|0\rangle$, the target state runs operation $S$. When the control state is $|1\rangle$, the target state runs operation $S^{\dag}$. When $t=M$, $W_{i}$ can be seen as the $i$th step of quantum walk with multiple coins.}
\label{fig:mcoins}
\end{figure}

In the above two quantum walks, there is only one coin space. Now we introduce quantum walks driven by many coins \cite{Brun_2003}. We take quantum walks on the line with many coins as an example, while quantum walks on graphs with many coins can be obtained in a similar way. This walk can be seen in Figure \ref{fig:mcoins}, where there are $M$ coins in total. The unitary transformation that results from flipping the $m$th coin is given by $W_m=(S\otimes P_{0m}+S^{\dag}\otimes P_{1m})(I\otimes C_{m})$. Here $C_{m}$ is a coin operator acting on the $m$th coin, $S=\sum_{n}|n+1\rangle|\langle n|$ is the shift operator on the position space,  and $P_{0m}=|0\rangle_{m}\langle0|$, $P_{1m}=|1\rangle_{m}\langle1|$. If we cycle among the coins, doing a total of $t$ flips ($t/M$ with each coin), then the state will be $|\Psi(t)\rangle=(W_M\cdot\cdot\cdot W _1)^{t/M}|\Psi_0\rangle$. Here we consider the situation where $t=M$. The number of the coin is the same as the steps of the walks. There are $M$ unitary transformations $\{W_{m}\}$, and they all commute with each other: $[W_{m},W_{n}]=0$ for $m\ne n$.


\section{III. Perfect state transfer by quantum walks with two coins}

\subsection{ Qubit state transfer on the line}
In the following, we will introduce a scheme to perfectly transfer the qubit (coin state) through quantum walks on one-dimensional line with two coins.
The initial state of the walker-coin-coin system is
\begin{equation}
|\phi\rangle^{0}=|0\rangle\otimes(a|0\rangle+b|1\rangle)\otimes|0\rangle,
\end{equation}
where $|a|^2+|b|^2=1$. Let $|\phi\rangle^{i}$ denote the state describing the system after the $i$th step.

In this scheme, the unknown state of coin 1 can be transferred to an arbitrary position $x$ after certain $n$ steps of quantum walks. The number of $n$ is correlated to $x$. The total process is described as
\begin{equation}
|\phi\rangle^{0} \stackrel{{n\mbox{\small { steps}}}}{\longrightarrow}  |\phi\rangle^{n}=|x\rangle\otimes U_1(a|0\rangle+b|1\rangle)\otimes U_2|0\rangle,
\label{pupose}
\end{equation}
where $U_1$ and $U_2$ are recovery unitary operators. Refer to Figure 2.

\begin{figure}[ht]
\centering
\includegraphics[width=0.5\linewidth]{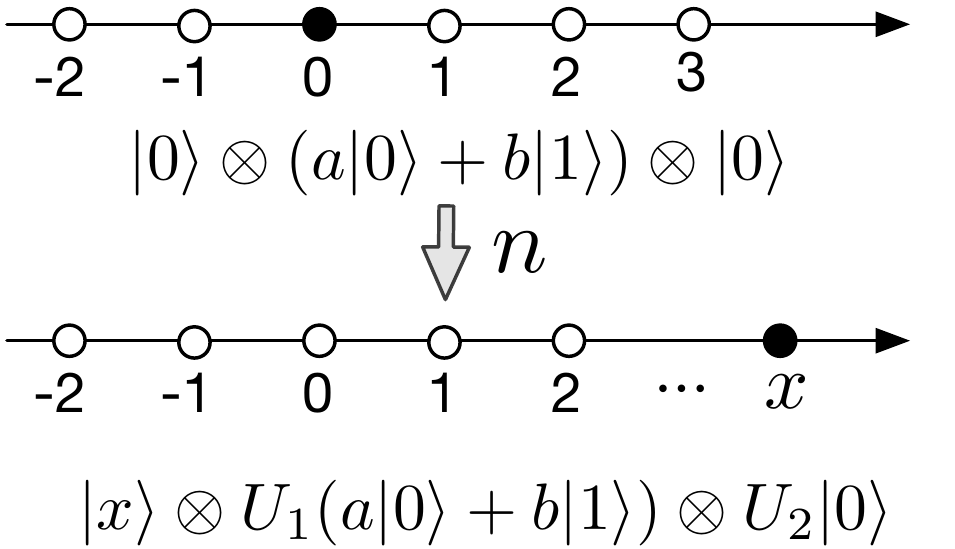}
\caption{Qubit state transfer on the one-dimensional line.
In this graph, the position $x$ is located at the right side of 0.
In addition, it can be in the left side.
The process is listed at the following cases.
}

\label{fig:1}
\end{figure}

It should be noted that we alternately use two coin operator $I$ and $X$.
At the odd (even) steps, we use the first (second) coin to control the movement of the walker.
If the coin operators are independent with the step $i$, the state of $|\phi\rangle^{i}$ is given by the following.
For an even number $i$, $|\phi\rangle^{i}=W_1^{i/2}W_2^{i/2}|\phi\rangle^{0}$.
For an odd number $i$, $|\phi\rangle^{i}=W_1^{\lceil{i/2}\rceil+1}W_2^{\lceil{i/2}\rceil}|\phi\rangle^{0}$.
In this scheme, the coin operators are related with the step $i$.
Here we list the selected coin operators under each step.
At the end of steps, we can choose suitable unitary operators $U_1$ and $U_2$ on the coin 1 and coin 2 space to recover the state.

\emph{Case 1.1:} The position $x$ is a positive and even number.

\begin{table}[ht]
\centering\scriptsize{
\begin{tabular}{|c|c|c|c|c|c|c|c|c|c|c|c|}
\hline
$i$th step & \ 1 \   & \  2 \   & $\cdots$      & $x-1$  & \ $x$ \     &$x+1$  &$x+2$ &$\cdots$   & $2x-1$  &\ $2x$ \  \\
\hline
$C_1$      &\ $I$ \  &          &$\cdots$       &$I$     &  \ \        & $X$   &      & $\cdots$  & $I$     & \ \ \\
\hline
$C_2$      &         &\ $I$ \   &$\cdots$       &        &\ $I$ \      &       & $I$  &$\cdots$   &         &\ $I$ \ \\
\hline

\end{tabular}}
\caption{\label{+2}{Case 1.1 on the line}}
\end{table}

In this case, only at the $(x+1)$-th step of quantum walks, we choose the coin operator $C_1$ on coin 1 space to be Pauli operator $X$.
At the other steps, the coin operators are all identity.
After we run $2x$ steps of quantum walks, the state $a|0\rangle+b|1\rangle$ can be perfectly transferred. The detailed process are given as follows£º
\begin{itemize}
\setlength{\itemsep}{0pt}
\item If $i=1$,
$
|\phi\rangle^{1}=|1\rangle\otimes a|00\rangle+|-1\rangle\otimes b|10\rangle.
$

\item If $i=2$,
$
|\phi\rangle^{2}=|2\rangle\otimes a|00\rangle+|0\rangle\otimes b|10\rangle.
$

\item If $i=x-1$,
$
|\phi\rangle^{x-1}=|x-1\rangle\otimes a|00\rangle+|-1\rangle\otimes b|10\rangle.
$

\item If $i=x$,
$
|\phi\rangle^{x}=|x\rangle\otimes a|00\rangle+|0\rangle\otimes b|10\rangle.
$
\end{itemize}

Here, we can check the situation that we run $2t$ steps of quantum walks when $C_1=C_2=I$.
The walker will move $2t$ steps forward at the coin state $|00\rangle$,
$2t$ steps backward at the coin state $|11\rangle$, or stand still at the coin state $|10\rangle$ ($|01\rangle$).

After $x$ (even) steps of quantum walks totally, the information of $a$ flows to the position $x$. And the information of $b$ rest on the position $0$.
In the rest steps of quantum walks, we should make the information of $b$ flow to the position $x$, and keep the information of $a$ still at the end step of quantum walks.
So we use the Pauli operator $X$ to entirely flip the coin states in the coin 1 space (i.e. $|0\rangle \rightarrow |1\rangle$, $|1\rangle \rightarrow |0\rangle$).
Then the coin state in the two coin spaces changes from $|00\rangle$ to $|10\rangle$ and from $|10\rangle$ to $|00\rangle$.
This will bring the result we need.

\begin{itemize}
\setlength{\itemsep}{0pt}
\item If $i=x+1$,
$
|\phi\rangle^{x+1}=|x-1\rangle\otimes a|10\rangle+|1\rangle\otimes b|00\rangle.
$

\item If $i=x+2$,
$
|\phi\rangle^{x+2}=|x\rangle\otimes a|10\rangle+|2\rangle\otimes b|00\rangle.
$

\item If $i=2x-1$,
$
|\phi\rangle^{2x-1}=|x-1\rangle\otimes a|10\rangle+|x-1\rangle\otimes b|00\rangle.
$

\item If $i=2x$,
$
|\phi\rangle^{2x}=|x\rangle\otimes a|10\rangle+|x\rangle\otimes b|00\rangle.
$
\end{itemize}

The final state $|\phi\rangle^{2x}$ is equal to $|x\rangle\otimes (a|1\rangle +b|0\rangle)\otimes|0\rangle$.
We can choose the recovery operator $U_1$ to be $X$. Then the state is successfully transferred to the position $x$.

We notice that the perfectly state transfer can be also achieved at the $(2x-1)$-th steps of quantum walk.
and the position $2x-1$ is odd. The positive and odd case is bellow.

\emph{Case 1.2:} The position $x$ is a positive and odd number.

\begin{table}[ht]
\centering\scriptsize{
\begin{tabular}{|c|c|c|c|c|c|c|c|c|c|c|c|c|}
\hline
$i$th step & \ 1 \   & \  2 \   & $\cdots$      & \  $x$ \  & \ $x+1$ \     &$x+2$  &$x+3$ &$\cdots$   & $2x-1$  &$2x$  &$2x+1$\\
\hline
$C_1$      &\ $I$ \  &          &$\cdots$       &\ $I$ \    &  \ \        & $X$   &      & $\cdots$  & $I$     &      &$I$\\
\hline
$C_2$      &         &\ $I$ \   &$\cdots$       &           &\ $I$ \      &       & $I$  &$\cdots$   &         &$I$   &\\
\hline

\end{tabular}}
\caption{\label{+1}{Case 1.2 on the line}}
\end{table}

In this case, only at the $(x+2)$-th step, we choose the coin operator $C_1$ on coin 1 space to be Pauli operator $X$.
At the other steps, the coin operators are all identity. We can perfectly transferred the unknown state with $2x+1$ steps. $U_1=X$, $U_2=I$.
The proof is similar with the even case.

\emph{Case 1.3:} The position $x$ is a negative and even number.

\begin{table}[ht]
\centering\scriptsize{
\begin{tabular}{|c|c|c|c|c|c|c|c|c|c|c|c|}
\hline
$i$th step & \ 1 \   & \  2 \   & $\cdots$      & $|x|-1$  & \ $|x|$ \     &$|x|+1$  &$|x|+2$ &$\cdots$   & $|2x|-1$  &\ $|2x|$ \  \\
\hline
$C_1$      &\ $I$ \  &          &$\cdots$       &$I$     &  \ \        & $X$   &      & $\cdots$  & $I$     & \ \ \\
\hline
$C_2$      &         &\ $X$ \   &$\cdots$       &        &\ $I$ \      &       & $I$  &$\cdots$   &         &\ $I$ \ \\
\hline
\end{tabular}}
\caption{\label{-2}{Case 1.3 on the line}}
\end{table}

In this case, we choose the coin operator $C_2$ to be $X$ at the second step of quantum walks, and we choose $C_1$ to be $X$ at the $(x+1)-th $ step.
At the other steps, the coin operators are all identity.
After we run $2|x|$ steps of quantum walks, where $x$ is a negative number, the state $a|0\rangle+b|1\rangle$ can be perfectly transferred.
The recovery operators are given by : $U_1=X$, and $U_2=I$. The results can be calculated as follows:

\begin{itemize}
\setlength{\itemsep}{0pt}

\item If $i=1$,
$
|\phi\rangle^{1}=|1\rangle\otimes a|00\rangle+|-1\rangle\otimes b|10\rangle.
$

\item If $i=2$,
$
|\phi\rangle^{2}=|0\rangle\otimes a|01\rangle+|-2\rangle\otimes b|11\rangle.
$

\item If $i=|x|-1$,
$
|\phi\rangle^{|x|-1}=|1\rangle\otimes a|01\rangle+|1-|x|\rangle\otimes b|11\rangle.
$

\item If $i=|x|$,
$
|\phi\rangle^{|x|}=|0\rangle\otimes a|01\rangle+|x\rangle\otimes b|11\rangle.
$

\end{itemize}

Here, the information of $b$ flows to the position $x$ and the information of $a$ rests on the position $0$.
\begin{itemize}
\setlength{\itemsep}{0pt}
\item If $i=|x|+1$,
$
|\phi\rangle^{|x|+1}=|-1\rangle\otimes a|11\rangle+|1-|x|\rangle\otimes b|01\rangle.
$

\item If $i=|x|+2$,
$
|\phi\rangle^{|x|+2}=|-2\rangle\otimes a|11\rangle+|x\rangle\otimes b|01\rangle.
$

\item If $i=2|x|-1$,
$
|\phi\rangle^{2|x|-1}=|1-|x|\rangle\otimes a|11\rangle+|1-|x|\rangle\otimes b|01\rangle.
$

\item If $i=2|x|$,
$
|\phi\rangle^{2|x|}=|x\rangle\otimes a|11\rangle+|x\rangle\otimes b|01\rangle.
$
\end{itemize}

The final state $|\phi\rangle^{2|x|}$ is equal to $|x\rangle\otimes (a|1\rangle +b|0\rangle)\otimes|1\rangle$.
Let $U_1=U_2=X$. Then the state is successfully transferred to the position $x$.

Also we can notice that the perfectly state transfer can be also achieved at the $(2|x|-1)$-th steps of quantum walk.
And the position $1-|2x|$ is odd. The negative and odd case is bellow.

\emph{Case 1.4:} The position $x$ is a negative and odd number.

\begin{table}[ht]
\centering \scriptsize{
\tabcolsep 0.003in
\begin{tabular}{|c|c|c|c|c|c|c|c|c|c|c|c|}
\hline
$i$th step &\ 1 \ & \  2 \ & $\cdots$ & \  $|x|$ \  & \ $|x|+1$ \  &$|x|+2$  &$|x|+3$ &$\cdots$   & $|2x|-1$  &$|2x|$  &$|2x|+1$\\
\hline
$C_1$      &\ $I$ \ &  &$\cdots$ &\ $I$ \ &  \ \  & $X$ & & $\cdots$ & $I$ & &$I$\\
\hline
$C_2$      &  &\ $X$ \  &$\cdots$ &  &\ $I$ \ &  & $I$  &$\cdots$ & &$I$ &\\
\hline

\end{tabular}}
\caption{\label{-1}{Case 1.4 on the line}}
\end{table}

In this case, we choose the coin operator $C_2$ to be $X$ at the second step of quantum walks and we choose $C_1$ to be $X$ at the $(|x|+2)-th $ step.
At the other steps, the coin operators are all identity.
After we run $2|x|+1$ steps of quantum walks, the state $a|0\rangle+b|1\rangle$ can be perfectly transferred.
The recovery operators are given by $U_1=X$, and $U_2=X$. The proof is similar with the even case.

\emph{Summery:} In this scheme, the coin operators are just dependent on the evolution of the time. They are independent with the positions.
The coin operators are selected from $\{I,X\}$.
There are at most two moments that we choose the coin operator to be $X$.

The process has a periodicity.
We take the case that $x$ is an even number as an example.
 In this process, we do not use the recovery operators.
Let the state $|\phi\rangle^{0}=|0\rangle\otimes (a|0\rangle +b|1\rangle)\otimes|0\rangle$ be the initial state.
After we run the process given in case 1.1, the final state is
$|\phi\rangle^{2x}=|x\rangle\otimes (a|1\rangle +b|0\rangle)\otimes|0\rangle$.
Let it be the initial state and run a new round of process given in case 1.3, and then the final state is
$|\phi\rangle^{4x}=|0\rangle\otimes (a|0\rangle +b|1\rangle)\otimes|0\rangle$.
This state returns to the initial state of the system.

Moreover, the scheme can be used to perfectly transfer the unknown coin state (qubit) from the initial position to any target position.
Let $l$ be the initial position and $|l\rangle\otimes(a|0\rangle +b|1\rangle)\otimes|0\rangle$ be the initial state of quantum walks.
If we run the process given in the first two scheme, the final state will be $|l+x\rangle\otimes(a|0\rangle +b|1\rangle)\otimes|0\rangle$.
The first two schemes can be used to transfer the state towards right with $|x|$ steps, i.e., from an initial position $l$ to $l+|x|$.
The last two schemes can be used to transfer the state towards left with $|x|$ steps, i.e., from $l$ to $l-|x|$.
Thus the schemes can be used as an efficient quantum routing.

\subsection{Qubit state transfer on circles}
In the schemes above, we use quantum walks on the line.
In this subsection, we can use quantum walks on graphs \cite{Aharonov_2001}.
Let $G(V,E)$ be a graph, and $\mathcal{H}_{v}$ be the Hilbert space spanned by states $|v\rangle$, where vertex $v\in V$.
At each vertex $j$, there are several directed edges with labels pointing to the other vertices.
The coin space $\mathcal{H}_c$ is spanned by states $|a\rangle$, where $a\in\{0,\cdots,m\}$.
They are the labels of the directed edges.
The conditional shift operation between $\mathcal{H}_v$ and $\mathcal{H}_c$ is
$T=\sum_{j,a}|k\rangle\langle j|\otimes|a\rangle\langle a|$, where the label $a$ directs the edge $j$ to $k$.

Choose a special graph, i.e., a circle with $d$ vertices, where there are two edges at each vertex. Then the coin space is spanned by $|0\rangle,|1\rangle$. The conditional shift operator between the position space and the coin space is defined as
\begin{equation}
\begin{split}
S_1=\sum_{k=0}^{d-1}(|(k+1)\bmod\,d\rangle\langle k|\otimes|0\rangle\langle0|+\\
|(k-1)\bmod\,d\rangle\langle k|\otimes|1\rangle\langle1|)
\end{split}
\end{equation}

In the paper \cite{Yalcinkaya_2015}, perfect state transfer can be achieved by quantum walks on $N$-circles with one coin space, where there are even vertices on the circle.
The unknown state is transferred from the initial position to the opposite position
i.e. $0\rightarrow N/2$, $N$ is an even number. However, in our schemes,
by introducing a new coin space, perfect state transfer can be achieved on circles with any vertices and from the initial position to any target position in our schemes.

\begin{figure}[ht]
\centering
\includegraphics[width=0.7\linewidth]{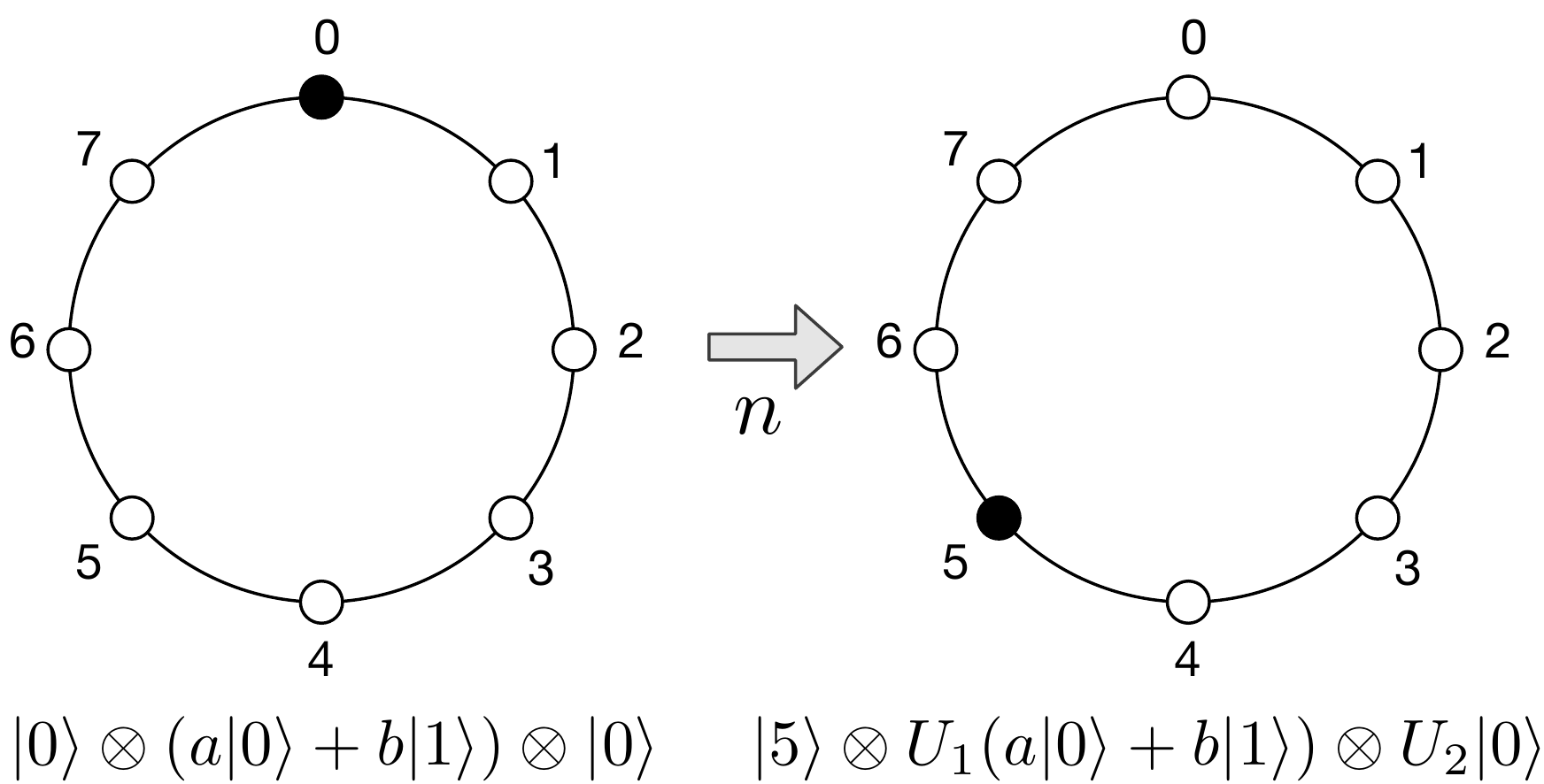}
\caption{Qubit state transfer on circles}\label{fig:2}
\end{figure}

Unlike with quantum walks on the line, it holds that $0\le x\le d-1$ for a certain vertex $x$ on the circle.
The whole process of transferring is described in equation \ref{pupose}, where $0\le x\le d-1$.
There is only one direction from the initial position $0$ to the target position $x$ in quantum walks on the line.
But there are two directions in quantum walks on circle, namely, clockwise and anticlockwise.
Here, we can check the situation that we ran $2$ steps of quantum walks when $C_1=C_2=I$.
The walker on the circle will rotate $2$ steps clockwise at the coin state $|00\rangle$,
$2$ steps anticlockwise at the coin state $|11\rangle$, or stand still at the coin state $|10\rangle$ ($|01\rangle$).
Thus, for transferring the state from 0 to $x$, there are four methods to design the scheme:
\begin{enumerate}
\setlength{\itemsep}{0pt}
\item The information of $a$ and $b$ flow from $0$ to $x$ clockwise.

\item The information of $a$ and $b$ flow from $0$ to $x$ anticlockwise.

\item The information of $a$ ($b$) flows from $0$ to $x$ clockwise (anticlockwise).

\item The information of $a$ ($b$) flows from $0$ to $x$ anticlockwise (clockwise).

\end{enumerate}

Now we introduce the first method.
The number of vertex $d$ can be odd or even.
If the vertex $x$ is even, we use the coin operators shown in Table \ref{+2}.
If the vertex $x$ is odd, we use the coin operators show in Table \ref{+1}.

The conditional shift operators in quantum walks on the line and circle are not identical.
However, after the calculation about the evolutions of quantum walks and unitary recovery,
the final states are same, both of which are $|x\rangle\otimes(a|0\rangle+b|1\rangle)\otimes|0\rangle$.
In this method, the total steps of quantum walks are $2x$ (even case) or $2x+1$ (odd case), respectively.

If the vertex $x$ is closer to 0 in anticlockwise direction than in clockwise direction, (i.e., $x>d/2$), we can use the second method for state transfer.
The total steps of quantum walks will be less than the first method. In addition, the number of vertices can be even or odd.
The total steps of quantum walks are $2d-2x$ (even case) or $2d-2x+1$ (odd case), respectively.

\emph{Case 2.1:} The number $d-x$ is even.
\begin{table}[ht]
\centering \scriptsize{
\tabcolsep 0.00009in
\begin{tabular}{|c|c|c|c|c|c|c|c|c|c|c|c|}
\hline
$i$th step & \ 1 \   & \  2 \   & $\cdots$      & $d-x-1$   & $d-x$    &$d-x+1$  &$d-x+2$ &$\cdots$   & $2d-2x-1$  &$2d-2x$  \\
\hline
$C_1$      &\ $I$ \  &          &$\cdots$       &$I$     &  \ \        & $X$   &      & $\cdots$  & $I$     & \ \ \\
\hline
$C_2$      &         &\ $X$ \   &$\cdots$       &        &\ $I$ \      &       & $I$  &$\cdots$   &         &\ $I$ \ \\
\hline
\end{tabular}}
\caption{\label{--2}{Case 2.1 on the circle}}
\end{table}

\emph{Case 2.2:} The number $d-x$ is odd.

\begin{table}[ht]
\centering\tiny{
\tabcolsep 0.00000009in
\begin{tabular}{|c|c|c|c|c|c|c|c|c|c|c|c|c|c|}
\hline
$i$th step & \ 1 \   & \  2 \   & $\cdots$      &  $d-x$   &  $d-x+1$     &$d-x+2$  &$d-x+3$ &$\cdots$   & $2d-2x-1$  &$2d-2x$  &$2d-2x+1$\\
\hline
$C_1$      &\ $I$ \  &          &$\cdots$       &\ $I$ \    &  \ \        & $X$   &      & $\cdots$  & $I$     &      &$I$\\
\hline
$C_2$      &         &\ $X$ \   &$\cdots$       &           &\ $I$ \      &       & $I$  &$\cdots$   &         &$I$   &\\
\hline

\end{tabular}}
\caption{\label{-1}{Case 2.2 on the circle}}
\end{table}


The last two methods can also be used to perfectly transfer a state.
The total steps of quantum walks are irrelevant with the number of $x$,
whereas the number of vertices should be even. We take the third method as an example, and then we list the coin operators at each step and the recovery unitary operators.

\emph{Case 3.1:} The position $x$ is even and the total number of vertices $d$ is even.

\begin{table}[ht]
\centering\scriptsize{
\begin{tabular}{|c|c|c|c|c|c|c|c|c|c|c|c|}
\hline
$i$th step & \ 1 \   & \  2 \   & $\cdots$      & $x-1$  & \ $x$ \     &$x+1$  &$x+2$ &$\cdots$   & $d-1$  &\ $d$ \  \\
\hline
$C_1$      &\ $I$ \  &          &$\cdots$       &$I$     &  \ \        & $I$   &      & $\cdots$  & $I$     & \ \ \\
\hline
$C_2$      &         &\ $I$ \   &$\cdots$       &        &\ $I$ \      &       & $X$  &$\cdots$   &         &\ $I$ \ \\
\hline

\end{tabular}}
\caption{\label{++2}{Case 3.1 on the circle}}
\end{table}

In this case, only at the $(x+2)$-th step, we let $C_2=X$. At the other steps, the coin operators are all identity. The calculation process is as follows£º

\begin{itemize}
\setlength{\itemsep}{0pt}
\item If $i=1$,
$
|\phi\rangle^{1}=|1\rangle\otimes a|00\rangle+|d-1\rangle\otimes b|10\rangle.
$

\item If $i=2$,
$
|\phi\rangle^{2}=|2\rangle\otimes a|00\rangle+|0\rangle\otimes b|10\rangle.
$

\item If $i=x-1$,
$
|\phi\rangle^{x-1}=|x-1\rangle\otimes a|00\rangle+|d-1\rangle\otimes b|10\rangle.
$

\item If $i=x$,
$
|\phi\rangle^{x}=|x\rangle\otimes a|00\rangle+|0\rangle\otimes b|10\rangle.
$
\end{itemize}

After $x$ steps of quantum walks totally, the information of $a$ flows to the position $x$. Moreover, the information of $b$ rests on the position $0$.
In the rest steps of quantum walks, we should make the information of $b$ flow to the position $x$ and keep the information of $a$ still at the last step.

\begin{itemize}
\setlength{\itemsep}{0pt}
\item If $i=x+1$,
$
|\phi\rangle^{x+1}=|x+1\rangle\otimes a|00\rangle+|d-1\rangle\otimes b|10\rangle.
$

\item If $i=x+2$,
$
|\phi\rangle^{x+2}=|x\rangle\otimes a|01\rangle+|d-2\rangle\otimes b|11\rangle.
$

\item If $i=x+(d-x-1)$,
$
|\phi\rangle^{d-1}=|x+1\rangle\otimes a|01\rangle+|x+1\rangle\otimes b|11\rangle.
$

\item If $i=d$,
$
|\phi\rangle^{d}=|x\rangle\otimes a|01\rangle+|x\rangle\otimes b|11\rangle.
$
\end{itemize}

The final state $|\phi\rangle^{d}$ is equal to $|x\rangle\otimes (a|0\rangle +b|1\rangle)\otimes|1\rangle$, which means that
the state is successfully transferred to the position $x$. Let $U_2=X$.

We notice that the perfectly state transfer can be also achieved at the $(2x-1)$-th steps of quantum walk, and
the position $2x-1$ is odd. The positive and odd case is bellow.

\emph{Case 3.2:} The position $x$ is odd and the total number of vertices $d$ is even.

\begin{table}[ht]
\centering\scriptsize{
\begin{tabular}{|c|c|c|c|c|c|c|c|c|c|c|c|c|}
\hline
$i$th step & \ 1 \   & \  2 \   & $\cdots$      &   $x-2$   & \ $x-1$ \     &\ $x$ \   &$x+1$ &$\cdots$   & $d-1$  & \ $d$ \   &$d+1$\\
\hline
$C_1$      &\ $I$ \  &          &$\cdots$       & $I$     &  \ \        & \ $I$ \  &      & $\cdots$  & $I$     &      &$I$\\
\hline
$C_2$      &         &\ $I$ \   &$\cdots$       &           &\ $I$ \      &       & $X$  &$\cdots$   &         &\ $I$ \    &\\
\hline

\end{tabular}}
\caption{\label{++1}{Case 3.2 on the circle}}
\end{table}
In this case, only at the $(x+1)$-th step, let $C_1=I$. At the other steps, the coin operators are all identity.
The proof is similar with the even case.

%
%
%
%
%
%
%
%
%
%

This process of walking on the circle also has a periodicity.
The total process can be described as that the initial state rotates $x$ steps clockwise from the initial position.
If we rotate the new state $d-x$ steps clockwise, the state will return to the initial position.

\textbf{}

\subsection{Qudit state transfer on the complete graph}

The scheme above is to transfer a qubit $a|0\rangle+b|1\rangle$ by quantum walks either on the line or on the circle.
The information of $a$ and $b$ are transferred from position 0 to $x$.
How about to be if we want to transfer the information of $\{a_0,\cdots,a_{d-1}\}$?
Now we introduce a scheme to perfectly transfer the qudit (coin state) through quantum walks on the complete graph with two coins.
The qudit is $\sum_{k=0}^{d-1}a_k|k\rangle$, where $|a_k|^2=1$.

Choose a special graph, i.e., a complete graph with $d$ vertices and add a loop at each vertex.
Each coin space is spanned by $|0\rangle,\cdots,|d-1\rangle$.
The conditional shift operator between the position space and the coin space is defined as
\begin{equation}
S_2=\sum_{k=0}^{d-1}\sum_{j=0}^{d-1}|(k+j)\bmod\,d\rangle\langle k|\otimes|j\rangle\langle j|
\end{equation}
This operator simulates the shift regulations between the vertices under the different tossed coin states.

Prepare the initial state of the walker-coin-coin system to be
\begin{equation}
|\phi\rangle^{0}=|0\rangle\otimes(\sum_{k=0}^{d-1}a_k|k\rangle)\otimes|0\rangle,
\end{equation}
Let $|\phi\rangle^{i}$ denote the state describing the system after the $i$th step.

In this scheme, the qudit can be transferred to an arbitrary position $x$ (except $0$) after certain $n$ steps of quantum walks. In addition,
the number of $n$ is correlated to $x$. The total process is described as follows£º
\begin{equation}
|\phi\rangle^{0} \stackrel{{n\mbox{\small { steps}}}}{\longrightarrow}  |\phi\rangle^{n}=|x\rangle\otimes U_1(\sum_{k=0}^{d-1}a_k|k\rangle)\otimes U_2|0\rangle,
\end{equation}
where $U_1$ and $U_2$ are recovery unitary operators.

It should be noticed that the coin operations are qudit operation.
Define a permutation operation:
$$X_d=\sum_{j=0}^{d-1}|(j+1)\bmod\,d\rangle \langle j|, X_d |0\rangle=|1\rangle$$

 Here we list the coin operators at each step.
\begin{table}[ht]
\centering{\scriptsize
\tabcolsep 0.003in
\begin{tabular}{|c|c|c|c|c|c|c|c|c|c|c|c|}
\hline
$i$th step & \ 1 \   & \  2 \   & $\cdots$      & $2d-2x-1$   & $2d-2x$    &$2d-2x+1$  &$2d-2x+2$ &$\cdots$   & $2d-1$  &$2d$  \\
\hline
$C_1$      &\ $I$ \  &          &$\cdots$       &$I$     &  \ \        & $I$   &      & $\cdots$  & $I$     & \ \ \\
\hline
$C_2$      &         &\ $I$ \   &$\cdots$       &        &\ $I$ \      &       & $X_d$  &$\cdots$   &         &\ $I$ \ \\
\hline
\end{tabular}}
\caption{\label{--2}{Case on the complete graph}}
\end{table}

Only at the $(2d-2x+2)$-th step of quantum walks, let $C_2=X_d$.
At the other steps, the coin operators are all identity. When $C_1=C_2=I$, if we run two steps of quantum walks, the position will shift from 0 to $k$ at the coin state $|k\rangle|0\rangle$. The process can be calculated as follows£º
\begin{itemize}
 \item If $i=2d-2x$,
\begin{equation}
|\phi\rangle^{2d-2x}=\sum_{k=0}^{d-1}a_k|(d-x)k \bmod\,d\rangle|k\rangle|0\rangle
\end{equation}

\item If $i=2d-2x+1$,
\begin{equation}
|\phi\rangle^{2d-2x+1}=\sum_{k=0}^{d-1}a_k|[(d-x)k+k] \bmod\,d\rangle|k\rangle|0\rangle.
\end{equation}

\item If $i=2d-2x+2$,
\begin{equation}
|\phi\rangle^{2d-2x+2}=\sum_{k=0}^{d-1}a_k|[(d-x)k+(k+1)] \bmod\,d\rangle|k\rangle|1\rangle
\end{equation}

\item If $i=2d$,
\begin{equation}
|\phi\rangle^{2d}=\sum_{k=0}^{d-1}a_k|[(d-x)k+x(k+1)] \bmod\,d\rangle|k\rangle|1\rangle
\end{equation}
\end{itemize}

Because $[(d-x)k+x(k+1)] \bmod\,d=x$, the state is $|x\rangle\otimes (\sum_{k=0}^{d-1}a_k|k\rangle)\otimes |1\rangle$. Let $U_2=X_d^{-}$.
The qudit is perfectly transferred from 0 to $x$.

Remark: In this scheme, if the coin operators are all identity at all these $2d$ steps, the qudit can be revived at position 0.

\section*{Qudit state transfer on the d-regular graph}
Now we introduce a scheme to perfectly transfer the qudit (coin state) through quantum walks on the $d$-regular graph with two coins. Denote the number of vertices by $n$.
The qudit is $\sum_{k=0}^{d-1}a_k|k\rangle$, where $|a_k|^2=1$.
Each coin space is spanned by $|0\rangle,\cdots,|d-1\rangle$.
The conditional shift operator between the position space and the coin space is defined as
\begin{equation}
S=\sum_{k=0}^{n-1}\sum_{j=0}^{d-1}|(k+j)\bmod\,n\rangle\langle k|\otimes|j\rangle\langle j|
\end{equation}
This operator simulates the shift regulations between the vertices under the different tossed coin states.

Prepare the initial state of the walker-coin-coin system to be
\begin{equation}
|\phi\rangle^{0}=|0\rangle\otimes(\sum_{k=0}^{d-1}a_k|k\rangle)\otimes|0\rangle,
\end{equation}
Let $|\phi\rangle^{i}$ denote the state describing the system after the $i$th step.

In this scheme, the qudit can be transferred to an arbitrary position $x$ (except x=0) after certain $m$ steps of quantum walks.
However, the number of steps is correlated to $x$. The total process is described as
\begin{equation}
|\phi\rangle^{0} \stackrel{{m\mbox{\small { steps}}}}{\longrightarrow}  |\phi\rangle^{m}=|x\rangle\otimes U_1(\sum_{k=0}^{d-1}a_k|k\rangle)\otimes U_2|0\rangle,
\end{equation}
where $U_1$ and $U_2$ are recovery unitary operators.

It should be noted that the coin operations are qudit operation.
Define a permutation operation as follows£º

 $$X_n=\sum_{j=0}^{d-1}|(j+1)\bmod\,n\rangle \langle j|, X_n |0\rangle=|1\rangle$$
Here we list the coin operators at each step.
\begin{table}[ht]
\centering\scriptsize{
\tabcolsep 0.003in
\begin{tabular}{|c|c|c|c|c|c|c|c|c|c|c|c|}
\hline
$i$th step & \ 1 \   & \  2 \   & $\cdots$      & $2n-2x-1$   & $2n-2x$    &$2n-2x+1$  &$2n-2x+2$ &$\cdots$   & $2n-1$  &$2n$  \\
\hline
$C_1$      &\ $I$ \  &          &$\cdots$       &$I$     &  \ \        & $I$   &      & $\cdots$  & $I$     & \ \ \\
\hline
$C_2$      &         &\ $I$ \   &$\cdots$       &        &\ $I$ \      &       & $X_n$  &$\cdots$   &         &\ $I$ \ \\
\hline
\end{tabular}}
\caption{\label{--2}{Case on the $d$-regular graph}}
\end{table}

Only at the $(2n-2x+2)$-th step of quantum walks, we let $C_2=X_n$.
At the other steps, the coin operators are all identity.
If $C_1=C_2=I$, if we run two steps of quantum walks, the position will shift from 0 to $k$ at the coin state $|k\rangle|0\rangle$.
The detail process is as follows£º
\begin{itemize}
 \item If $i=2n-2x$,
\begin{equation}
|\phi\rangle^{2n-2x}=\sum_{k=0}^{d-1}a_k|(n-x)k \bmod\,n\rangle|k\rangle|0\rangle
\end{equation}

\item If $i=2n-2x+1$,
\begin{equation}
|\phi\rangle^{2n-2x+1}=\sum_{k=0}^{d-1}a_k|[(n-x)k+k] \bmod\,n\rangle|k\rangle|0\rangle.
\end{equation}

\item If $i=2n-2x+2$,
\begin{equation}
|\phi\rangle^{2n-2x+2}=\sum_{k=0}^{d-1}a_k|[(n-x)k+(k+1)] \bmod\,n\rangle|k\rangle|1\rangle
\end{equation}

\item If $i=2n$,
\begin{equation}
\begin{split}
|\phi\rangle^{2n}&=\sum_{k=0}^{d-1}a_k|[(n-x)k+x(k+1)] \bmod\,n\rangle|k\rangle|1\rangle
\\
&=|x\rangle\otimes (\sum_{k=0}^{d-1}a_k|k\rangle)\otimes |1\rangle
\end{split}
\end{equation}
\end{itemize}

Let $U_2=X_n^{-1}$.
The qudit is perfectly transferred from $0$ to $x$.

Remark: In this scheme, when the coin operators are all identity at all these $2n$ steps, the qudit can be revived at position $0$.

\section{IV. Generalized teleportation over long steps of walks by quantum walks with two coins}

 If the measurement is used as tools, the information can also be transferred between coins space. Recently, teleportation schemes by quantum walks with two coins has been introduced \cite{Wang_2017}. By two steps of quantum walks, we can transmit unknown qubit/qudit state from coin 1 space to coin 2 space. However, the teleportation based on quantum walks with two coins by long steps of walk remains unclear. In this section, we discuss the implementation conditions to conduct generalized teleportation by quantum walks with two coins on various models.

\subsection{Qubit teleportation on one-dimensional line over long steps of walk}

Suppose that there are two coins totally. The initial state of walker-coin-coin system is
\begin{equation}
|\phi\rangle^{0}=|0\rangle\otimes(a|0\rangle+b|1\rangle)\otimes|+\rangle.
\end{equation}

Our goal is to revive the state $a|0\rangle+b|1\rangle$ on coin 2 space in some positions.

In this scheme, local measurements in the position space and coin 1 space are allowed, and $S$ is chosen to be the evolution between position space and coin space. Let the coin operators to be identity, i.e., $C_1=C_2=I$.

After the first step of $W_1$, the state evolutes to
\begin{equation}
\begin{split}
|\phi\rangle^{1}&=(W_1^{1}W_2^{0})|\phi\rangle^{0}
\\
&=|1\rangle\otimes a|0\rangle\otimes|+\rangle+|-1\rangle\otimes b|1\rangle\otimes|+\rangle.
\end{split}
\end{equation}

After the second step of $W_2$, the state evolutes to
\begin{equation}
\begin{split}
|\phi\rangle^{2}&=(W_1^{1}W_2^{1})|\phi\rangle^{0}
\\
&=|0\rangle\otimes (a|01\rangle+ b|10\rangle)/\sqrt{2}+(|-2\rangle\otimes b|11\rangle
\\
&\qquad+|2\rangle\otimes a|00\rangle)/\sqrt{2}.
\end{split}
\end{equation}

Let $n$ be an even number. At time $t=n$,
 the state evolutes to
\begin{equation}
\begin{split}
|\phi\rangle^{t}&=(W_1^{n/2}W_2^{n/2})|\phi\rangle^{0}
\\
&=|0\rangle\otimes (a|01\rangle+ b|10\rangle)/\sqrt{2}
\\
&\qquad+(|-n\rangle\otimes b|11\rangle+|n\rangle\otimes a|00\rangle)/\sqrt{2}.
\end{split}
\end{equation}

Then, we take measurement with basis $\{|0\rangle,|\tilde{n}\rangle,|-\tilde{n}\rangle,\cdots\}$ on the position space and use the measurement basis $\{|+\rangle,|-\rangle \}$ on the coin 1 space, namely, $|\pm\tilde{n}\rangle=(|-n\rangle\pm|n\rangle)/\sqrt{2}$. Mark three outcomes $ -1,0,1$ corresponding to $|-\tilde{n}\rangle,|0\rangle,|\tilde{n}\rangle$ at the position space, respectively. In addition, mark outcomes $-1,1$ if the state at coin 1 space collapses to $|-\rangle,|+\rangle$, respectively. The state at coin 2 space will recover to the target state $a|0\rangle+b|1\rangle$ if operating some local unitary operations according to the measurement results. The table below lists the results and the corresponding revise operations.

\begin{table}[ht]
\centering
\begin{tabular}{|c|c|}
\hline
Measurement Results on position and coin 1 space& Revise operation \\
\hline
1,1  or -1,-1 & $I$  \\
\hline
-1,1 or 1,-1  & $Z$  \\
\hline
0,1           & $X$  \\
\hline
0,-1         & $ZX$  \\
\hline
\end{tabular}
\caption{\label{tab:qubit} Measurement results and revise operations}
\end{table}

\subsection{Qubit teleportation on circles over long steps of walk}
Suppose that there are two coins totally. The initial state of walker-coin-coin system is
\begin{equation}
|\phi\rangle^{0}=|0\rangle\otimes(a|0\rangle+b|1\rangle)\otimes|+\rangle.
\end{equation}
Let the coin operators be identity and independent with time, $C_1=C_2=I$.
After the first step of $W_1$, the state evolutes to
\begin{equation}
\begin{split}
|\phi\rangle^{1}&=(W_1^{1}W_2^{0})|\phi\rangle^{0}
\\
&=|1\rangle\otimes a|0\rangle\otimes|+\rangle+|d-1\rangle\otimes b|1\rangle\otimes|+\rangle.
\end{split}
\end{equation}
After the second step of $W_2$, the state evolutes to
\begin{equation}
\begin{split}
|\phi\rangle^{2}&=(W_1^{1}W_2^{1}|\phi\rangle^{0}
\\
&=|0\rangle\otimes (a|01\rangle+ b|10\rangle)/\sqrt{2}
\\
&\qquad+(|d-2\rangle\otimes b|11\rangle+|2\rangle\otimes a|00\rangle)/\sqrt{2}.
\end{split}
\end{equation}
Let the number of vertices $d$ be an even number. At time $t=d/2$,
 the state evolutes to

\begin{equation}
\begin{split}
|\phi\rangle^{t}&=(W_1^{d/4}W_2^{d/4})|\phi\rangle^{0}
\\
&=|0\rangle\otimes (a|01\rangle+ b|10\rangle)/\sqrt{2}
\\
&\qquad+|d/2\rangle\otimes (a|00\rangle+b|11\rangle)/\sqrt{2}.
\end{split}
\end{equation}

Then we take measurement with basis $\{|0\rangle,\cdots,|d-1\rangle\}$ on the position space and use the measurement basis $\{|+\rangle,|-\rangle \}$ on the coin 1 space. Mark two outcomes $0,1$ corresponding to $|0\rangle,|d/2\rangle$ at the position space, respectively. In addition, mark outcomes $-1,1$ if the state at coin 1 apace collapses to $|-\rangle,|+\rangle$, respectively. The state at coin 2 space will recover to the target state $a|0\rangle+b|1\rangle$ if operating some local unitary operations according to the measurement results. The table below lists the results and corresponding revise operation. In this scheme, we can revive the unknown state in two positions, by $1/2$ probability, respectively.
\begin{table}[ht]
\centering
\begin{tabular}{|c|c|}
\hline
Measurement Results on position and coin 1 space& Revise operation \\
\hline
1,1   & $I$  \\
\hline
1,-1  & $Z$  \\
\hline
0,1           & $X$  \\
\hline
0,-1         & $ZX$  \\
\hline
\end{tabular}
\caption{\label{tab:qubit} Measurement results and revise operations}
\end{table}


\subsection{Qudit teleportation on complete graph over long steps of walk}
In this scheme, after $2t$ steps, we can use quantum walks on $d$-regular graph to transfer an unknown $d$-dimensional qudit $\sum_{k=0}^{d-1}a_k|k\rangle$ from the first coin space to the second coin space, $\sum_{k=0}^{d-1}|a_k|^2=1$.
Choose a complete graph with $d$ vertices and add a loop at each vertex. The coin space is spanned by $|0\rangle,\cdots,|d-1\rangle$.
The conditional shift operator between the position space and the coin space is defined by
\begin{equation}
S_2=\sum_{k=0}^{d-1}\sum_{j=0}^{d-1}|(k+j)\bmod\,d\rangle\langle k|\otimes|j\rangle\langle j|
\end{equation}

Let the coin operator to be identity, $C_1=C_2=I$.
The initial state of walker-coin-coin system is
\begin{equation}
|\phi\rangle^{0}=|0\rangle\otimes\sum_{k=0}^{d-1}a_k|k\rangle\otimes \sum_{j=0}^{d-1}|j\rangle/\sqrt{d}.
\end{equation}
After the first step of $W_1$, the state evolutes to
\begin{equation}
\begin{split}
|\phi\rangle^{1}&=(W_1^{1}W_2^{0}|\phi\rangle^{0}
\\
&=\sum_{k=0}^{d-1}a_k|k\rangle\otimes|k\rangle\otimes \sum_{j=0}^{d-1}|j\rangle/\sqrt{d}.
\end{split}
\end{equation}
After the second step of $W_2$, the state evolutes to
\begin{equation}
\begin{split}
|\phi\rangle^{2}&=(W_1^{1}W_2^{1}|\phi\rangle^{0}
\\
&=\sum_{k=0}^{d-1}\sum_{j=0}^{d-1}a_{k}|(k+j)\bmod\,d\rangle|k\rangle|j\rangle/\sqrt{d}
\end{split}
\end{equation}
Let $t$ be an integer mutually prime with the dimension $d$. At time $2t$,
 the state evolutes to

\begin{equation}
\begin{split}
|\phi\rangle^{2t}&=(W_1^{t}W_2^{t}|\phi\rangle^{0}
\\
&=\sum_{k=0}^{d-1}\sum_{j=0}^{d-1}a_{k}|t\cdot(k+j)\bmod\,d\rangle|k\rangle|j\rangle/\sqrt{d}.
\end{split}
\end{equation}
Let $(k+j)\bmod\,d=s$, this state is equal to
\begin{equation}
\sum_{k=0}^{d-1}\sum_{s=0}^{d-1}a_{k}|t\cdot s\bmod\,d\rangle|k\rangle|(s-k)\bmod\,d\rangle/\sqrt{d}.
\end{equation}

Denote $(t \cdot s\bmod\,d)$ to be $x_s$. Certainly, $x_s\in\{0,\cdots,d-1\}$. For two different numbers $s_1$ and $s_2\in\{0,\cdots,d-1\}$, we have the relation $x_{s_1}\ne x_{s_2}$. Assume $x_{s_1}= x_{s_2}$, we conclude that $t\cdot (s_1-s_2)\bmod\,d=0$. Because $s_1,s_2<d$, $(s_1-s_2)\bmod\,d\ne0$, we have $t \bmod\,d=0$. This is contradictory with that $t$ and $d$ are mutually prime.

Then, we take measurement with basis $\{|k\rangle:k=0,\cdots,d-1\}$ on the position space. Mark the outcome with $k$ if the state collapse to $|k\rangle$. In addition, we take measurement with basis $\{|\tilde{k}\rangle:k=0,\cdots,d-1\}$ on the coin 1 space, which is obtained by Fourier transform:
\begin{equation}
|\tilde{k}\rangle=\sum\nolimits_{m=0}^{d-1}e^{2\pi imk/d}|m\rangle/\sqrt{d}, \quad k=0,1,\cdots,d-1.
\end{equation}
If we obtain result $x_s$ at the position space, the state in coin 1 and coin 2 space will evolute to

\begin{equation}
\sum_{k=0}^{d-1}a_{k}|k\rangle|(s-k)\bmod\,d\rangle.
\end{equation}
Correspondingly, this state can be expressed by
\begin{equation}
\sum_{\tilde{t}=0}^{d-1}[|\tilde{t}\rangle\otimes \sum_{k=0}^{d-1}a_{k}e^{-2\pi i\tilde{t}k/d}|(s-k)\bmod\,d\rangle]/\sqrt{d}.
\end{equation}
If the state at the coin 1 space collapses to $|\tilde{t}\rangle$, the state at the coin 2 space will be $\sum_{k=0}^{d-1}a_{k}e^{-2\pi i\tilde{t}k/d}|(s-k)\bmod\,d\rangle]$. Then state of coin 2 operates with unitary operation $U_{x_s\tilde{t}}$, given by
\begin{equation}
U_{x_s\tilde{t}}=\sum_{k=0}^{d-1}e^{2\pi ik\tilde{t}/d}|k\rangle\langle (s-k)\bmod\, d|.
\end{equation}
Thus, the state at coin 2 space will recover to $\sum_{k=0}^{d-1}a_k|k\rangle$.

\subsection{Qudit teleportation on $d$-regular graph over long steps of walk }
We can also select a $d$-regular graph with $n$ vertices to teleport an unknown qudit, where $n\ge 2d-1$. Suppose the number of vertices is in the range of $\{d+1,\cdots,2d-2\}$.
The qudit can not be teleported with probability 1. Let $ \lceil n/2d-1\rceil$ be the integer part, where $t\in\{1,\cdots,\lceil n/2d-1\rceil\}$. The teleportation scheme will occur at $2t$ steps. When $t>\lceil n/2d-1\rceil$, the existing state in the position space exceeds the maximum label of vertices.

The first step of the walk is given by $W_1=E_1\cdot(I_v\otimes C_1\otimes I_2)$, where $E_1=\sum_{j=0}^{d-1}\sum_{k=0}^{n-1}|(k+j)\bmod n\rangle\langle k|\otimes|j\rangle_1\langle j|\otimes I_{2}$. Similarly, the coin operator $C_1$ can be any qudit operation for successfully teleporting. Here let $C_1=I$.
 The second step of the walk is given by $W_2=E_2\cdot(I_v\otimes I_1\otimes I),$  where $E_2=\sum_{j=0}^{d-1}\sum_{k=0}^{n-1}|(k+j)\bmod n\rangle\langle k|\otimes I_{1}\otimes|j\rangle_2\langle j|$.
The initial state is
\begin{equation}
|0\rangle\otimes\sum\nolimits_{m=0}^{d-1}a_{m}|m\rangle\otimes \sum\nolimits_{k=0}^{d-1}|k\rangle/\sqrt{d}.
\end{equation}

After two steps of quantum walks, we have the state given by
\begin{equation}
\sum\nolimits_{m=0}^{d-1}\sum\nolimits_{k=0}^{d-1}a_{m}|m+k\rangle|m\rangle|k\rangle/\sqrt{d}.
\end{equation}

After $2t$ steps of walk, the final state is
\begin{equation}
\sum\nolimits_{m=0}^{d-1}\sum\nolimits_{k=0}^{d-1}a_{m}|t(m+k)\bmod\,n\rangle|m\rangle|k\rangle/\sqrt{d}.
\end{equation}

The set of all integers that mutually prime with $n$ and no more than $n$ is denoted by $A(n)$. Let $\phi(n)$ represent the size of $A(n)$.
That is to say,
\begin{equation}
\begin{split}
A(n)& =\{x|1\leq x\leq n, x\in Z^{+}, (x, n)=1\}
\\
&=\{A_1, A_2, \cdots, A_{\phi(n)}\}
\\
& = \{1, A_2, \cdots, A_{\phi(n)}\}
\end{split}
\end{equation}

For example, $A(6)=\{1,5\}, \phi(6)=2$; $A(9)=\{1,2,4,5,7,8\}, \phi(9)=6$.

Let $t=xn+A_i$, where $x=0, 1, 2, \cdots $, and $i=1, 2, \cdots, \phi(n)$.
So the final state is
$$|\phi\rangle^{2t}=\sum\nolimits_{m=0}^{d-1}\sum\nolimits_{k=0}^{d-1}a_{m}|A_{i}(m+k)\bmod\,n\rangle|m\rangle|k\rangle/\sqrt{d}.$$
Obviously, $(A_{i}(m+k)\bmod\,n)\in\{0,1,\cdots,n-1\}$.
 For arbitrary different values of $(m+k)$, they correspond to different $(A_{i}(m+k)\bmod\,n)$. There are totally $(2d-1)$ different states in the position space.
For convenience, we list the position state following the order of $(m+k)$ from small to large:
$$|0^{(i)}\rangle, |1^{(i)}\rangle, \cdots, |(2d-1)^{(i)}\rangle.$$

Thus, Alice measures the position state with basis

\begin{equation}
\begin{split}
\{(|k^{(i)}\rangle\pm|(d+k)^{(i)}\rangle)/\sqrt{2},|(d-1)^{(i)}\rangle,\cdots,|(n-1)^{(i)}\rangle\}
\label{basisd-regular}
\end{split}
\end{equation}
for $k=0,\cdots,d-2$.
Denote the result to be $k$, $k+d$ or $d-1$ if the state degenerates to be $(|k^{(i)}\rangle+|(d+k)^{(i)}\rangle)/\sqrt{2}$, $(|k^{(i)}\rangle-|(d+k)^{(i)}\rangle)/\sqrt{2}$ or $|(d-1)^{(i)}\rangle$, respectively.

 In addition, Alice measures coin 1 state with the basis $\{|\tilde{k}\rangle:k=0,\cdots,d-1\}$ obtained from Fourier transform. For example, when the result on the position space is $k$, the state of coin 1 and coin 2 is
\begin{equation}
\sum\nolimits_{m=0}^{k}a_m|m\rangle\otimes|k-m\rangle+\sum\nolimits_{m=k+1}^{d-1}a_m|m\rangle\otimes|k+d-m\rangle
\end{equation}
Expressing the state in coin 1 space with the basis obtained from Fourier transform, this state is equal to

\begin{equation}
\begin{split}
&\sum\nolimits_{\tilde{t}=0}^{d-1}(\sum\nolimits_{m=0}^{k}a_m e^{-2\pi i\tilde{t}m/d}|\tilde{t}\rangle\otimes|k-m\rangle+
\\
& \qquad\qquad\sum\nolimits_{m=k+1}^{d-1}a_m e^{-2\pi i\tilde{t}m/d}|\tilde{t}\rangle\otimes|k+d-m\rangle)
\end{split}
\end{equation}

If the result of coin 1 state is $\tilde{t}$, the state of coin 2 is
\begin{equation}
\begin{split}
&\sum\nolimits_{m=0}^{k}a_m e^{-2\pi i\tilde{t}m/d}|k-m\rangle+
\\&\qquad\qquad\sum\nolimits_{m=k+1}^{d-1}a_m e^{-2\pi i\tilde{t}m/d}|k+d-m\rangle.
\end{split}
\end{equation}
In order to recover the state $\sum_{m=0}^{d-1}a_m|m\rangle$, Bob performs local unitary operation represented by
\begin{equation}
\begin{split}
U_{k\tilde{t}}&=\sum\nolimits_{m=0}^{k}e^{2\pi i\tilde{t}m/d}|m\rangle\langle k-m|+
\\
&\qquad\sum\nolimits_{m=k+1}^{d-1}e^{2\pi i\tilde{t}m/d}|m\rangle\langle k+d-m|.
\end{split}
\end{equation}
If the results of the position state and coin 1 state are $d+k$ and $\tilde{t}$, Bob performs local unitary operation given by
\begin{equation}
\begin{split}
U_{(d+k),\tilde{t}}&=\sum\nolimits_{m=0}^{k}e^{2\pi i\tilde{t}m/d}|m\rangle\langle k-m|-
\\
&\qquad\sum\nolimits_{m=k+1}^{d-1}e^{2\pi i\tilde{t}m/d}|m\rangle\langle k+d-m|.
\end{split}
\end{equation}
If the results of the position state and coin 1 state is $d-1$ and $\tilde{t}$, respectively. Bob performs local unitary operation as follows:
\begin{equation}
U_{(d-1),\tilde{t}}=\sum\nolimits_{m=0}^{d-1}e^{2\pi i\tilde{t}m/d}|m\rangle\langle d-1-m|.
\end{equation}

\section*{V. Summary}
In this paper, we study quantum communication protocols on various building blocks for quantum networks by quantum walks with two coins. By alternately using two coins, we find that perfect qubit state transfer can be implemented either on the line or on the $N$-circle. In particular, perfect qudit state transfer by quantum walks can be first realized either on the complete graphs or on the $d$-regular graphs which cannot be obtained by quantum walks with one coin. Furthermore, we study how to do generalized teleportation \cite{Wang_2017} by quantum walk with two coins after long steps of walk on the above graphs. These works explore some new application for quantum walks with multiple coins. Because quantum walks are an universal quantum computing model, it may provide an universal platform for the implementation of scalable quantum networks.

\section{Acknowldgement}
This work was partially supported by National Key Research and Development Program of China under grant 2016YFB1000902, National Natural Science Foundation of China (Grant No.61472412), and Program for Creative Research Group of National Natural Science Foundation of China (Grant No. 61621003).

\end{document}